# A new measure of phase synchronization for a pair of time series and seizure focus localization

Kaushik Majumdar, Institute of Mathematical Sciences, Chennai – 600113, India. E-mail: kaushik@imsc.res.in

**Abstract**

Defining and measuring phase synchronization in a pair of nonlinear time series are highly nontrivial. This can be done with the help of Fourier transform, when it exists, for a pair of stored (hence stationary) signals. In a time series instantaneous phase is often defined with the help of Hilbert transform. In this paper phase of a time series has been defined with the help of Fourier transform. This gives rise to a deterministic method to detect phase synchronization in its most general form between a pair of time series. Since this is a stricter method than the statistical methods based on instantaneous phase, this can be used for lateralization and source localization of epileptic seizures with greater accuracy. Based on this method a novel measure of phase synchronization, called $syn$ function, has been defined, which is capable of quantifying neural phase synchronization and asynchronization as important parameters of epileptic seizure dynamics. It has been shown that such a strict measure of phase synchronization has potential application in seizure focus localization from scalp electroencephalogram (EEG) data, without any knowledge of electrical conductivity of the head.

## Introduction

History of synchronization dates back to Huygens [1]. In seventeenth century he studied the motion of two pendulums when they oscillated simultaneously in the same way, which is called synchronization. When two time series or trajectories change their course in same way, even at different time, they may be thought as phase synchronous. More generally, if they do not change in the same way, but maintain a constant ratio in course of their alteration, they can still be thought as phase synchronized or phase locked or phase coupled. Two dynamical systems whose trajectories maintain this property are phase synchronized.

The effect of phase synchronization of chaotic systems has been described in [2], [3]. It is synchronization of periodic oscillations, where only the phase locking is important, while no restriction on the amplitudes is imposed. Amplitudes are chaotic and may remain quite uncorrelated [4]. Of course the very notions of phase and amplitude of chaotic systems are rather nontrivial. Phase synchronization has found applications in laser dynamics [5], [6], solid state physics [7], electronics [8], biology [9] and communication [10].

Phase synchronization is an important phenomenon in neuroscience for both physiological and pathological reasons. Information processing in the mammalian brains takes place over distributed areas in the cerebral cortex and a possible candidate for binding those processing together to form a single unified perception is synchronous firing of a group of neurons [11]. During epileptic seizures the human brain experiences synchronous firing by an unusually large number of neurons, which can be detected from scalp EEG. Proper detection and analysis of synchronization among the EEG time series (which are chaotic in nature) of a patient, collected through strategically placed electrodes, is an important objective of epilepsy research [12], [13]. Detection of an appropriately defined synchronization between two nonlinear time series is a key to successful development of seizure prediction and control tools [14].

Both in medicine and physics there is no unified definition of synchronization. A notion of phase synchronization as developed in [24] has been used to study spatiotemporal changes in the pattern of synchronization in scalp EEG signals to have a deeper insight into the dynamics of epileptic seizure [23]. Spatiotemporal synchronization and asynchronization in EEG signals carry important brain dynamical information related to epileptic seizures [15], which at least in some cases may properly be utilized with the help of a stricter measure of synchronization. Most studies report that visual inspection of scalp EEG signals leads to 80–95% of correct side detection for epileptic focus [15]. A tight measure of phase synchronization has good potentiality to improve this study and to narrow down the epileptogenic zone (EZ) closer to the seizure focus.

## 2. Method

$$m\alpha - n\beta = C, \tag{1}$$

where $C$ is constant, $\alpha, \beta$ are two different phases belonging to two different trajectories and $m, n$ are integers. (1) gives the most general definition of *phase synchronization*, some time referred to as $m:n$ *phase locking* [23]. $m, n$ and $C$ are fixed, but unknown. Even finding $\alpha, \beta$ is highly nontrivial. When the equality of (1) is replaced by an inequality, a weaker condition, we get $m:n$ *phase entrainment* [16]. The notion of phase synchronization is best suited along a single frequency or at the most along a very narrow frequency band. In that case any synchronization detection method should be applied to a frequency window $f \pm \kappa$, $f$ is the frequency of interest and $\kappa$ is a small positive integer. This can be obtained by a suitable band pass filter. The method that is described below can obviously be applied in a windowed manner over the entire frequency range. No further discussion on windowing the frequency range will be taken up in this paper.

*A. Detection algorithm*

Let $x(t)$ be the output of a system $S_1$ and $y(t)$ be the output of another system $S_2$ and both are signal like, in the sense that they can be expressed as Fourier series, where the series of course converges point wise to the signal on a time interval $[0, T]$, where $T$ is the duration of observation or period. $x(t)$ and $y(t)$ can be expressed as in [17], i.e.,

$$x(t) = \frac{a_0^x}{2} + \sum_{i=1}^{\infty}(a_i^x \cos\frac{2\pi it}{T} + b_i^x \sin\frac{2\pi it}{T}) \tag{2}$$

$$a_i^x = \frac{2}{T}\int_0^T x(t)\cos\frac{2\pi it}{T}dt \tag{3}$$

$$b_i^x = \frac{2}{T}\int_0^T x(t)\sin\frac{2\pi it}{T}dt\,. \tag{4}$$

$$y(t) = \frac{a_0^y}{2} + \sum_{i=1}^{\infty}(a_i^y \cos\frac{2\pi it}{T} + b_i^y \sin\frac{2\pi it}{T}) \tag{5}$$

$$a_i^y = \frac{2}{T}\int_0^T y(t)\cos\frac{2\pi it}{T}dt \tag{6}$$

$$b_i^y = \frac{2}{T}\int_0^T y(t)\sin\frac{2\pi it}{T}dt\,. \tag{7}$$

$x(t)$ and $y(t)$ can be rewritten as

$$x(t) = \frac{a_0^x}{2} + \sum_{i=1}^{\infty} ((a_i^x)^2 + (b_i^x)^2)^{1/2} \sin\left( \frac{2\pi i t}{T} + \alpha_i^x \right) \quad (8)$$

$$\alpha_i^x = \tan^{-1}\left( \frac{a_i^x}{b_i^x} \right), \quad (9)$$

$$y(t) = \frac{a_0^y}{2} + \sum_{i=1}^{\infty} ((a_i^y)^2 + (b_i^y)^2)^{1/2} \sin\left( \frac{2\pi i t}{T} + \alpha_i^y \right) \quad (10)$$

$$\alpha_i^y = \tan^{-1}\left( \frac{a_i^y}{b_i^y} \right). \quad (11)$$

for $t \in [0,T]$. The phases of $x(t)$ and $y(t)$ are distributed over all harmonic components and they are given by $\sum_{i=1}^{\infty} \alpha_i^x \bmod 2\pi$ and $\sum_{i=1}^{\infty} \alpha_i^y \bmod 2\pi$ respectively. Unlike in $\tan^{-1}\left( \frac{\mathrm{H}(x(t))}{x(t)} \right)$, where $H(x(t))$ is the Hilbert transform of $x(t)$, which gives instantaneous phase of $x(t)$ at time $t$, provided $\mathrm{H}(x(t))$ exists, $\sum_{i=1}^{\infty} \alpha_i^x \bmod 2\pi$ gives a global phase of $x(t)$ whenever (2), (3) and (4) are valid. This is a direct generalization of the classical notion of phase difference in basic trigonometric functions, which should be valid for stored signals.

If $x(t)$ and $y(t)$ are phase synchronous then by (1), equation (12) will have to be true.

$$m\left( \sum_{i=1}^{\infty} \alpha_i^x \bmod 2\pi \right) - n\left( \sum_{i=1}^{\infty} \alpha_i^y \bmod 2\pi \right) = C', \quad (12)$$

where $C'$ is a constant. Clearly if

$$(m\alpha_i^x - n\alpha_i^y) \bmod 2\pi = (m\alpha_j^x - n\alpha_j^y) \bmod 2\pi = C \tag{13}$$

for all $i, j$ such that $i \neq j$, then (12) holds, where $C$ is a constant. So (13) is a sufficient condition for (12) to be true. Also for a stationary signal, where amplitudes are fixed and not varying with time any more the Fourier analysis given by (2) through (11) is valid. This means that the harmonic wise phase are given by (9) and (11). Any information regarding phase difference between $x(t)$ and $y(t)$ also lies in (9) and (11). The $ith$ harmonic component of $x(t)$ is phase synchronous with the $ith$ harmonic component of $y(t)$, because they have constant phase difference. If $x(t)$ and $y(t)$ are to be phase synchronous or almost phase synchronous the harmonic wise phase difference between the two signals must remain the same or almost the same over all the harmonics. This must also hold during the period irrespective of the rate of sample frequency. So (12) implies (13) i.e., (13) is also a necessary condition for (12) to hold for sampled stored signals. Notice that for each individual component $m\alpha_i^x - n\alpha_i^y \bmod 2\pi = m\alpha_i^x - n\alpha_i^y$ and therefore (13) can be rewritten as

$$m\alpha_i^x - n\alpha_i^y = m\alpha_j^x - n\alpha_j^y = C \tag{14}$$

for all $i, j$ such that $i \neq j$. So (14) is necessary and sufficient for two discrete signals $x(t)$ and $y(t)$ (sampled and digitized from the original continuous signals) to be in phase synchronization. Since to determine synchronization the signals must have to be subjected to computation, by taking digital signals in place of continuous ones no generality is lost.

(14) can be rewritten in matrix form as

$$\begin{bmatrix} \tan^{-1}\frac{a_i^x}{b_i^x} & -\tan^{-1}\frac{a_i^y}{b_i^y} \\ \tan^{-1}\frac{a_j^x}{b_j^x} & -\tan^{-1}\frac{a_j^y}{b_j^y} \end{bmatrix} \begin{bmatrix} z \\ w \end{bmatrix} = \begin{bmatrix} 1 \\ 1 \end{bmatrix}, \tag{15}$$

where $z = \frac{m}{C}$, $w = \frac{n}{C}$, $C \neq 0$ and $i \neq j$. (15) yields the following solution in determinant form:

$$z = \frac{\begin{vmatrix} 1 & -\tan^{-1}\frac{a_i^y}{b_i^y} \\ 1 & -\tan^{-1}\frac{a_j^y}{b_j^y} \end{vmatrix}}{\Delta} \tag{16}$$

$$w = \frac{\begin{vmatrix} \tan^{-1}\frac{a_i^x}{b_i^x} & 1 \\ \tan^{-1}\frac{a_j^x}{b_j^x} & 1 \end{vmatrix}}{\Delta}, \tag{17}$$

$$\Delta = \begin{vmatrix} \tan^{-1}\frac{a_i^x}{b_i^x} & -\tan^{-1}\frac{a_i^y}{b_i^y} \\ \tan^{-1}\frac{a_j^x}{b_j^x} & -\tan^{-1}\frac{a_j^y}{b_j^y} \end{vmatrix} \neq 0. \tag{18}$$

If (14) is to hold then $[z\ w]^T$ must be a fixed vector (in this particular case the superscript $T$

stands for transpose) and therefore all solutions obtained for (15) for different pairs of $i, j$, such that $i \neq j$, must be the same. In reality all calculations are error prone due to truncation of numbers and therefore in case of exact phase synchronization or close to exact phase synchronization the values of $z$ and $w$ as calculated by (16) through (18) should be convergent. This means difference between consecutive values of $z$ should converge to zero, with almost vanishing standard deviation. Same should be the case for $w$. How close a value to zero will be acceptable for the mean and standard deviation of the differences of consecutive values of $z$ and $w$ that will depend on the specific application. In any case a reasonable measure of synchronization between $x(t)$ and $y(t)$ can be defined as

$$syn(x(t), y(t)) = std(Z) + std(W), \tag{19}$$

where $syn$ can be called the *synchronization function* or in short $syn$ function, and $std$ stands for standard deviation. $syn$ function has been defined in a slightly different way in [20], where it has been taken $m = n = 1$ all along. $std(Z)$ is to be calculated over all the values of differences in consecutive values of $z$ obtained by replacing $j$ by $i+1$ in (16) and (18). Similarly $std(W)$ is to be calculated over all the values of differences between consecutive values of $w$ obtained by replacing $j$ by $i+1$ in (17) and (18).

When $\Delta \to 0$, it is clear from (16) and (17) that both $z$ and $w$ become arbitrarily large and therefore the difference between at least a pair of values from each of them also becomes arbitrarily large. In that case $std(Z) + std(W)$ becomes very large. This indicates that $x(t)$ and $y(t)$ are asynchronous in at least one harmonic and therefore they are asynchronous. So in the limiting case, when $\Delta = 0$, the signals are *completely asynchronous*.

The following computer algorithm will either return a value of the *syn* function or declare "Signals are asynchronous" if the signals are completely asynchronous. It is obvious that the algorithm will work even if (1) is replaced by $|m\alpha - n\beta - C| \le \varepsilon$, where $\varepsilon \ge 0$ is small.

**Proc(synchronization_detection)**

**Input:** $x(t), y(t), T$;

**Output:** $syn(x(t), y(t))$;

**Signals are asynchronous;** /* When they are completely asynchronous in some harmonic.*/

1. $A \leftarrow$ **FFT(** $x(t),[0, T]$**);** /* $A([0]) \leftarrow \frac{a_0^x}{2}$ */

2. $B \leftarrow$ **FFT(** $y(t),[0, T]$**);** /* $B([0]) \leftarrow \frac{a_0^y}{2}$ */

3. **for (** $2 \le i \le \left\lceil \frac{T}{2} \right\rceil - 1$ **)**

$$\Delta([i]\,[i+1]) \leftarrow \begin{vmatrix} \tan^{-1}\frac{\mathrm{Re}(A[i])}{\mathrm{Im}(A[i])} & -\tan^{-1}\frac{\mathrm{Re}(B[i])}{\mathrm{Im}(B[i])} \\ \tan^{-1}\frac{\mathrm{Re}(A[i+1])}{\mathrm{Im}(A[i+1])} & -\tan^{-1}\frac{\mathrm{Re}(B[i+1])}{\mathrm{Im}(B[i+1])} \end{vmatrix};$$

**If(** $\Delta([i]\,[i+1]) = 0$ **)**

**printf("Signals are asynchronous");**

**return;**

**else**

$$z([i]\,[i+1]) \leftarrow \frac{\begin{vmatrix} 1 & -\tan^{-1}\dfrac{\mathrm{Re}(B[i])}{\mathrm{Im}(B[i])} \\ 1 & -\tan^{-1}\dfrac{\mathrm{Re}(B[i+1])}{\mathrm{Im}(B[i+1])} \end{vmatrix}}{\Delta([i]\,[i+1])};$$

$$Z([i]) \leftarrow \left| z([i][i+1]) - z([i-1][i]) \right|;$$

$$w([i]\,[i+1]) \leftarrow \frac{\begin{vmatrix} \tan^{-1}\dfrac{\mathrm{Re}(A[i])}{\mathrm{Im}(A[i])} & 1 \\ \tan^{-1}\dfrac{\mathrm{Re}(A[i+1])}{\mathrm{Im}(A[i+1])} & 1 \end{vmatrix}}{\Delta([i]\,[i+1])};$$

$$W([i]) \leftarrow \left| w([i][i+1]) - w(([i-1][i]) \right|;$$

**end for;**

**4.** $std(Z)$;

**5.** $std(W)$;

**return(** $syn(x(t), y(t)) \leftarrow std(Z) + std(W)$ **);**

Each of steps 1 and 2 will take $O(T \log T)$ time [18], the for-loop of step 3 will take $O(T)$ time, each of step 4 and step 5 will take $O(T)$ time. So the total time taken by Proc(synchronization_detection) is $O(T \log T)$. In other words it is possible to say in $O(T \log T)$ time if two dynamical systems are in phase synchronization with each other, and if they are then to what extent as given by the value of the $syn$ function, provided their trajectories can be represented as Fourier series. Let us call such dynamical systems as *Fourier dynamics*.

*B. Application in seizure*

Temporal lobe epilepsy (TLE) is the most common form of partial or focal epilepsy. For about half the patients with TLE surgery is the only available therapy and the success of it depends on an accurate localization of the epileptozenic zone (EZ), defined as the brain area responsible for the generation of seizures. The first step towards a precise definition and localization of EZ is lateralization. Usually not one but several noninvasive multimodal approaches are employed for this purpose. The main idea is to finding the EZ by pair wise synchronization and asynchronization at theta band (4–8 Hz) among the scalp EEG signals as suggested in [15]. Correlation in theta band EEG is often observed at the seizure onset. A nonlinear regression analysis based amplitude correlation between pair wise scalp EEG channels on a particular side of the head is able to lateralize the EZ [15].

For a good localization seizure EEG waves shall have to be detected over a significant part of the head and therefore a high resolution sensor net such as in Fig. 1 will be useful for collecting as much EEG data as possible from a wide area of the head. However electrodes will have to be selected judiciously. The pair of electrodes not more than two centimeters apart will show significant synchronization due to volume conduction. Inputs from electrodes falling outside the scalp area (as in Fig. 1) will have to be eliminated from the study. When there is only one focus the localization problem will be less challenging. Let us assume this is the case. After necessary filtering of the data with a band pass filter appropriate frequency band is to be selected and artifacts will have to be removed.

A crucial stage in the process of automatic lateralization is to automatically identify the epoch of onset of seizure from the scalp EEG data. Various methods have been reviewed in [19]. A special case of the algorithm described above appears in [20] where it has been taken $m = n = 1$ and the $syn$ function is defined in a slightly different manner. This algorithm also takes $O(T \log T)$ time to execute to detect synchronization between two channels for the time duration $T$. If the data is collected from $C$ number of electrodes, then there are total of $\frac{1}{2}C(C-1)$ number of readings i.e., different values of $syn$ function. They can be written in a triangular matrix form as $syn_{ij}, 1 \le i < j \le C$. Decrease in pair wise synchronization at certain recording sites before an onset of seizure has been reported for a large number of cases in [21]. Pair wise synchronization in recording electrodes during seizure over the EZ has been reported in [15], [21]. The reason behind it is not very clearly understood, but it is almost obvious that the onset of a seizure is signified by a high degree of synchronization in a number of recording pairs. This is denoted by a transition in the matrix $[syn_{ij}]_T$ from large values at most of the entries to very small values to a significant number of entries, where $T$ is the duration of recording. $T$ may be suitably windowed. The time complexity of the whole operation is $O(CT(C+\log T))$ [20] and $T$ is long. The transition may be expressed as from $\sum_{\substack{i,j=1 \\ i<j}}^{C} syn_{ij} > \delta_2$ to $\sum_{\substack{i,j=1 \\ i<j}}^{C} syn_{ij} < \delta_1$, for some suitably chosen $\delta_1$ and $\delta_2$ such that $0 < \delta_1 < \delta_2$. The epoch, say $T_0$ ($0 < T_0 < T$), at which this transition happens can be taken to be the onset of seizure according to [21].

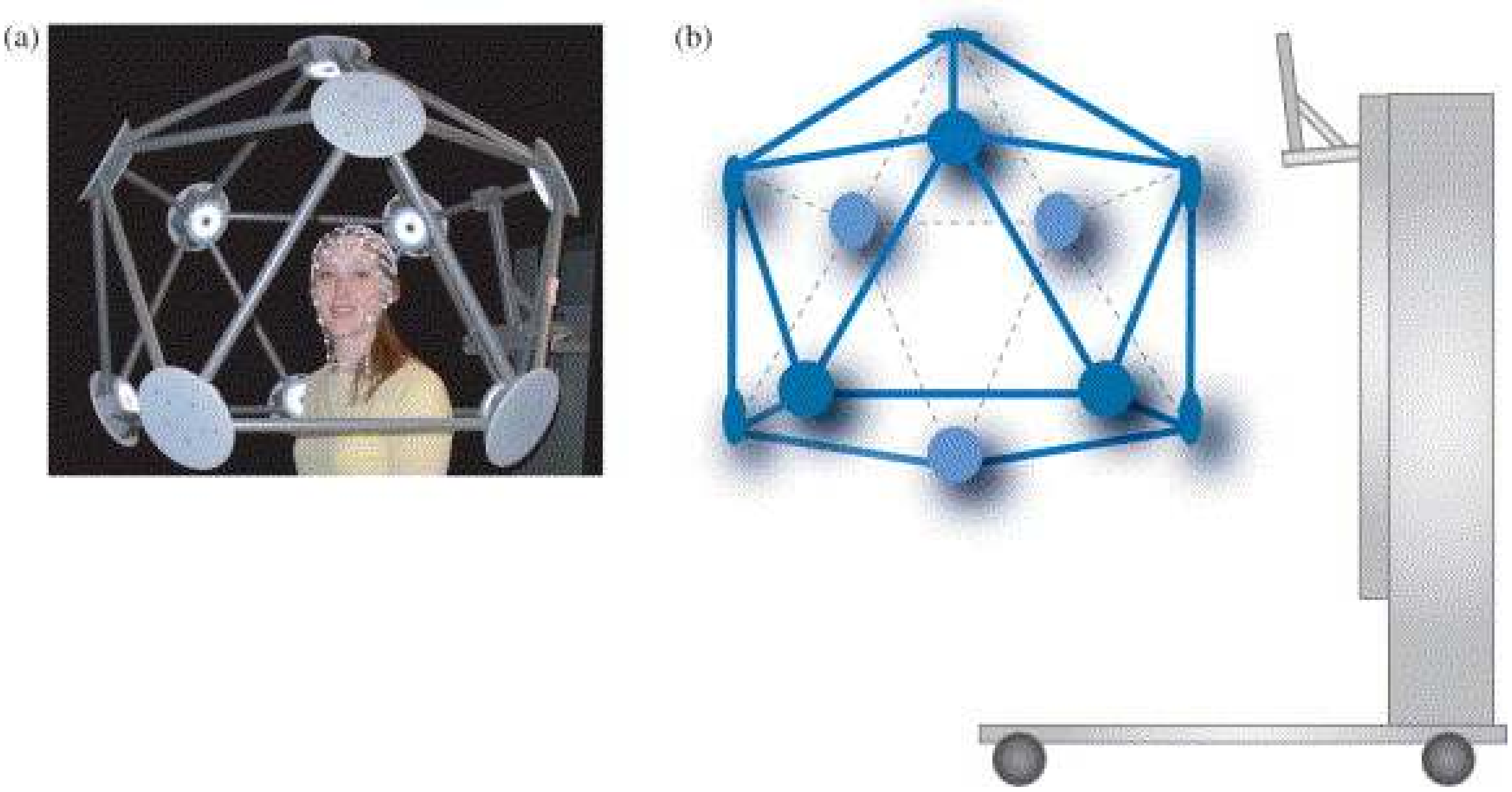

Fig. 1. Illustration of the geodesic photogrammetry system. (a) Subject under camera dome; and (b) camera dome and gantry. This dense array net covers all parts of the head with high spatial resolution and can be used for epileptic seizure focus localization with judicious choice of electrodes. The three dimensional coordinate system $S$ is to be fixed on the camera dome. Adopted from [22].

Next take the interval $[T_0, T_1]$, where $T_0 < T_1 < T$ and $T_1 - T_0$ is very small, only a couple of seconds. A good sample frequency, say 1000 Hz, of signals in $[T_0, T_1]$ will be very useful. The matrix $[syn_{ij}]_{T_1 - T_0}$ will have to be calculated. Note that each entry of the matrix gives synchronization value between a pair of electrodes, but to determine the source at least four electrodes are necessary which are pair wise synchronous with lowest value of $syn$ function. The idea is if these four points are not lying in the same plane they will be on a unique sphere and the source will be at the center of the sphere.

The four points can be selected automatically by a computer algorithm or may be selected manually by an expert by checking the entries of the matrix $[syn_{ij}]_{T_1 - T_0}$. The detail of the computer algorithm will be discussed in a separate paper. Once the group of four or more electrodes are identified, such that between any pair of them the value of the $syn$ function is close to zero, the next task is to pinpoint their location on the head with respect to a fixed spatial (Euclidean) coordinate system $S$. This can be done with photogrammetry method for dense array EEG channels as reported in [22]. 11 fixed positioned cameras in a camera dome is used for the purpose (Fig. 1, (b)) and three dimensional electrode positions are determined as shown in a two dimensional projection (Fig. 2).

Next step is to coordinatize the collection of points whose pair wise synchronization is very high. Each subset of four points will have to be tested for coplanarity according to the following determinant equation

$$\begin{vmatrix} x_1 & y_1 & z_1 & 1 \\ x_2 & y_2 & z_2 & 1 \\ x_3 & y_3 & z_3 & 1 \\ x_4 & y_4 & z_4 & 1 \end{vmatrix} \neq 0, \tag{20}$$

where $(x_1, y_1, z_1)$, $(x_2, y_2, z_2)$, $(x_3, y_3, z_3)$ and $(x_4, y_4, z_4)$ are the four points (coordinatized electrode positions). If (20) holds then the four points are not coplanar i.e., not in the same plane. Let the unique sphere passing through them has center at $(a, b, c)$ and a specific radius. Then the following equations hold.

$$a(x_1 - x_2) + b(y_1 - y_2) + c(z_1 - z_2) = \frac{1}{2}\left(x_1^2 + y_1^2 + z_1^2\right) - \frac{1}{2}\left(x_2^2 + y_2^2 + z_2^2\right), \tag{21}$$

$$a(x_2 - x_3) + b(y_2 - y_3) + c(z_2 - z_3) = \frac{1}{2}\left(x_2^2 + y_2^2 + z_2^2\right) - \frac{1}{2}\left(x_3^2 + y_3^2 + z_3^2\right), \tag{22}$$

$$a(x_3 - x_1) + b(y_3 - y_1) + c(z_3 - z_1) = \frac{1}{2}\left(x_3^2 + y_3^2 + z_3^2\right) - \frac{1}{2}\left(x_1^2 + y_1^2 + z_1^2\right). \tag{23}$$

Solving (21), (22) and (23) the position $(a, b, c)$ with respect to $S$ is determined, which is the location of the seizure focus or the source of synchronous EEG signals immediately after the onset of seizure.

If there are $D$ number of synchronous EEG channels for the duration of $[T_0, T_1]$ then there will be at most ${}^{D}C_4$ number of different source locations. If there is only one focus and the measurement error is insignificant, all these points should lie within a small neighborhood, which must contain the seizure focus.

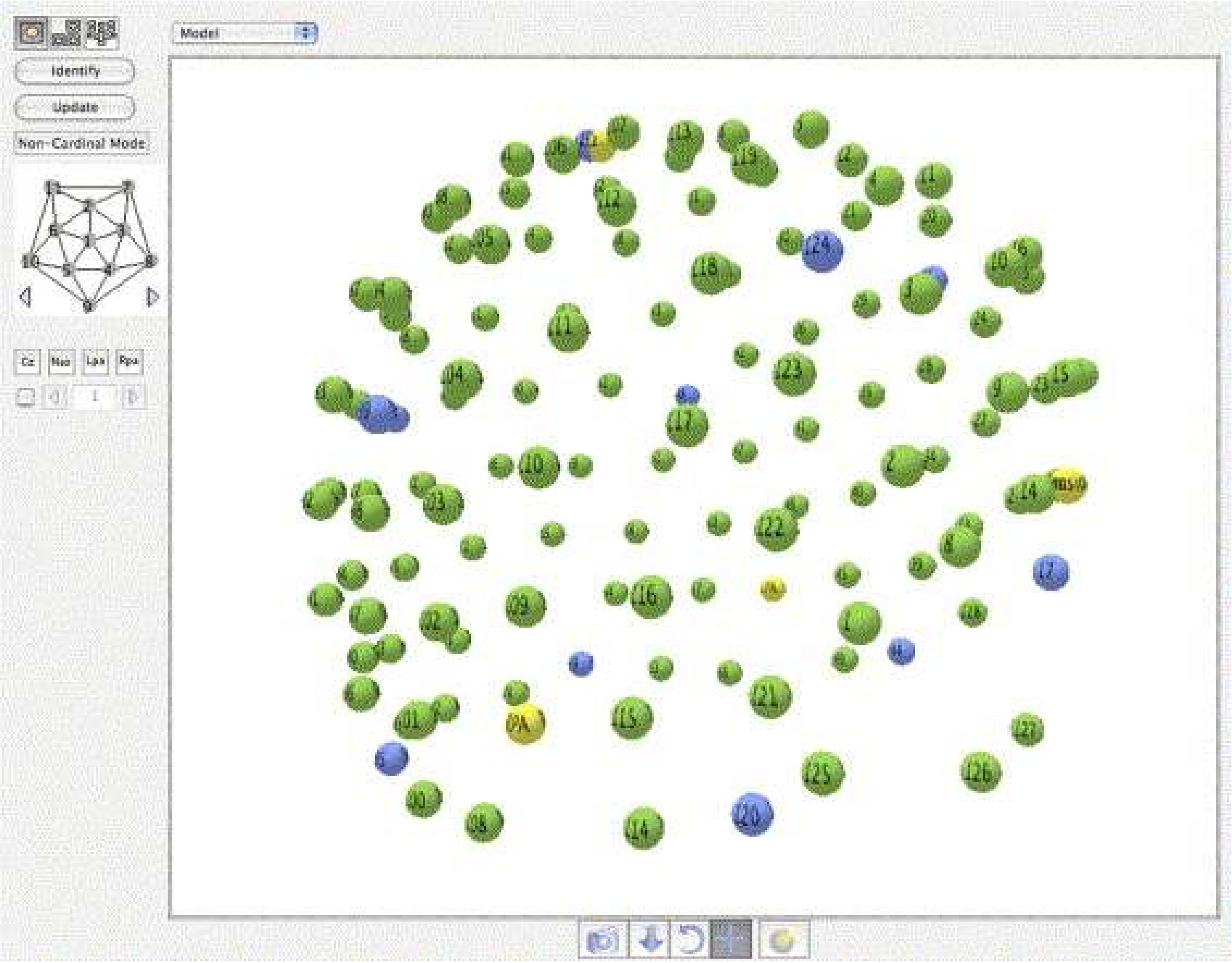

Fig. 2. Side view of solved three dimensional sensor positions. Adopted from [22].

### 3. Results

It is obvious that when electric field from a source inside the skull is propagating outwards to the scalp it may have different amplitude at different points on the scalp depending on the amount of attenuation suffered due to conductivity of the head tissue. This conductivity is different at different places and also at different angle of incidence. Since head is not spherical the internal electric field generated from any focal source will penetrate the skull at different angles at different points. It will suffer least attenuation when penetrating the skull normally. However as long as the field is coming from the same source when it will be detected on the scalp at different channels their phase will remain the same. In other words the signals detected at different channels will be phase locked with each other despite having different amplitude distributions. In this section the algorithm will be tested on four synthetic, almost phase locked signals with significant variation in amplitude (Fig. 3). Here $m = n = 1$ and therefore a special version of the algorithm described in [20] for the $m = n = 1$ case will be implemented. The four EEG like signals in Fig. 3 be denoted by $x_1(t), x_2(t), x_3(t)$ and $x_4(t)$. They have been marked as 1, 2, 3 and 4 respectively in the figure.

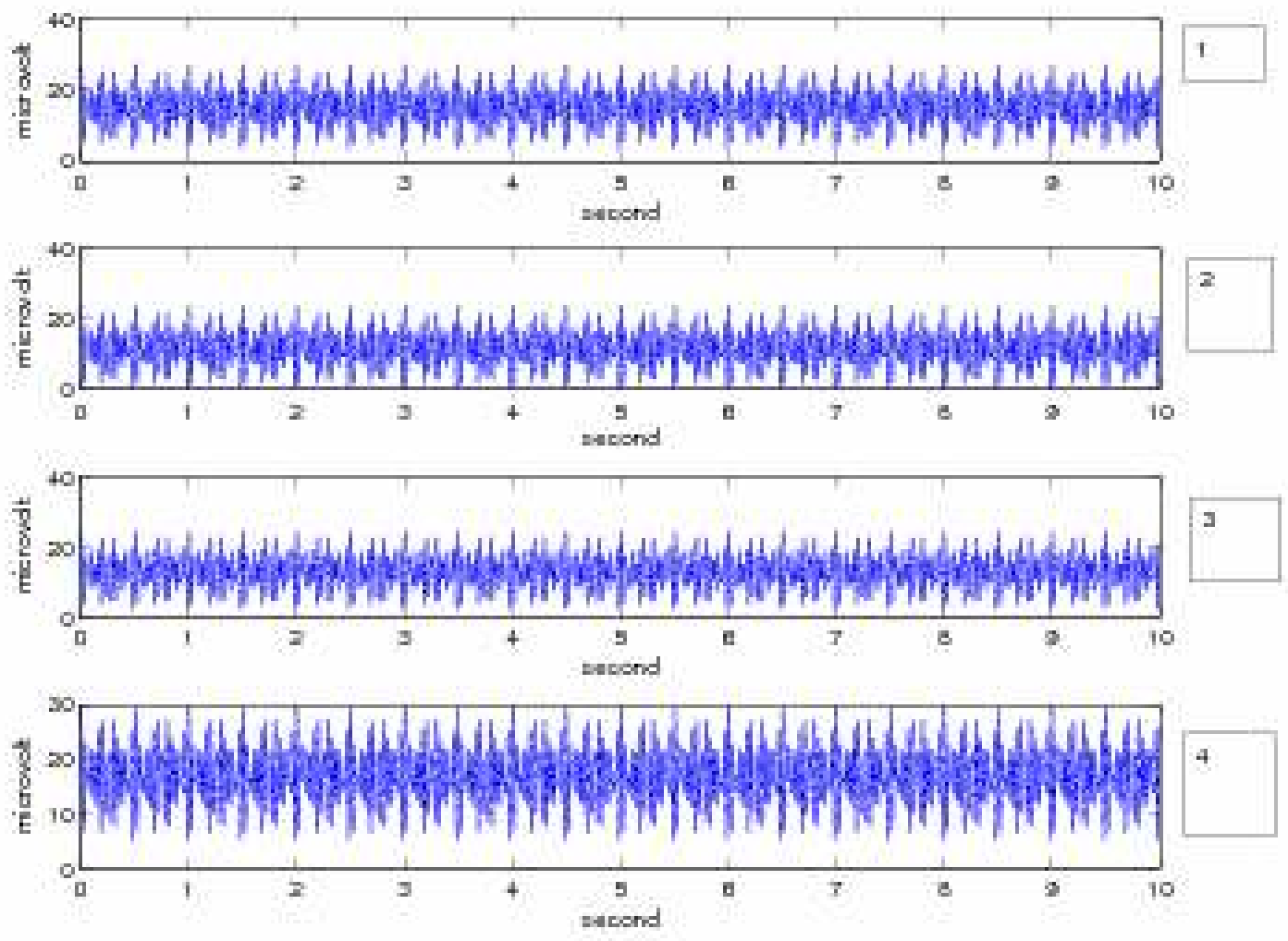

Fig. 3. Four signals in phase synchronization, although their amplitudes are quite different. They have been generated by (24), (25), (26) and (27).

$$x_1(t) = 15 + 2.75\cos(40\pi t) + 2.8105\sin(40\pi t) + 3.19\cos(48\pi t) + 2.8545\sin(48\pi t) + 3.025\cos(60\pi t) + 2.805\sin(60\pi t)\ . \tag{24}$$

$$x_2(t) = 11.5 + 2.9095\cos(40\pi t) + 2.75\sin(40\pi t) + 3.025\cos(48\pi t) + 2.805\sin(48\pi t) + 2.75\cos(60\pi t) + 2.86\sin(60\pi t)\ . \tag{25}$$

$$x_3(t) = 13 + 3.2005\cos(40\pi t) + 3.025\sin(40\pi t) + 2.42\cos(48\pi t) + 2.244\sin(48\pi t) + 2.613\cos(60\pi t) + 2.717\sin(60\pi t)\ . \tag{26}$$

$$x_4(t) = 17 + 2.42\cos(40\pi t) + 2.4732\sin(40\pi t) + 3.4452\cos(48\pi t) + 3.0229\sin(48\pi t) + 3.267\cos(60\pi t) + 3.029\sin(60\pi t)\ . \tag{27}$$

The four signals are synchronous, because $x_1(t)$ and $x_2(t)$ are as in [20], except for the constant terms and it has been shown that they are phase synchronous. $x_3(t)$ been derived from $x_2(t)$ by varying the constant term and then varying the coefficients of $nth$ harmonic in such a way that $\frac{a_n^3}{b_n^3} = \frac{a_n^2}{b_n^2}$ holds for all $n$, where superscript 3 corresponds to signal 3 in Fig. 3, etc. Exactly the same way $x_4(t)$ has been derived from $x_1(t)$. The results are as follows.

Table 1. Values of pair wise $syn$ functions of the signals in Fig. 3.

| | |
|---|---|
| $syn(x_1(t), x_2(t))$ | 0.0087 |
| $syn(x_2(t), x_3(t))$ | 0.5779 |
| $syn(x_3(t), x_4(t))$ | 0.3288 |
| $syn(x_4(t), x_1(t))$ | 0.0160 |
| $syn(x_1(t), x_3(t))$ | 0.6017 |
| $syn(x_2(t), x_4(t))$ | 0.0853 |

The signals have been sampled at 1000 Hz. All the calculations have been done using MATLAB. The $syn$ function is given by the version of the algorithm implemented in [20]. $x_3(t)$ seems to be rather asynchronous with the other signals. This is because of accumulation of more round off errors at the time of constructing $x_3(t)$ from $x_2(t)$. This shows the algorithm's sensitivity to even small amount of asynchrony. Noise is said to cause phase slip [16], which may make synchronous signals asynchronous. To cope with this a greater threshold of the $syn$ function value can be fixed to determine asynchronization. This may show some asynchronous signals as synchronous. However it will be able to identify all closely synchronous signals unambiguously.

## 4. Conclusion

Since the present method gives a sharper and stricter criterion for phase synchronization it can be utilized to classify the dynamical systems whose trajectories are representable by Fourier series. Two such systems are phase synchronous if and only if for a trajectory $x_i^1(t)$ in the system $S_1$ there is a unique trajectory $x_i^2(t)$ in the system $S_2$ and vice versa, such that $x_i^1(t)$ and $x_i^2(t)$ are phase synchronous in the sense of (1). This relation is an equivalence relation, for it is reflexive and symmetric. To prove the transitivity let us observe if three trajectories from $S_1$, $S_2$ and $S_3$ obey $m\alpha - n\beta = C_1$ and $p\beta - q\gamma = C_2$ then $pm\alpha - qn\gamma = pC_1 + nC_2$. $\alpha$, $\beta$ and $\gamma$ are the phases of the trajectories. Clearly the class of Fourier dynamics can be decomposed into equivalence classes. In pure mathematics equivalence classes are disjoint. Here a compromise can be reached in the sense class boundaries will overlap depending on the thresholding value chosen for the *syn* function.

Focal seizure may not necessarily originate from a pinpointed focus, but from a wider cortical network [23]. The method described in this paper will work best when there is only one pinpointed seizure focus not too deep inside the brain. When there is a spread out network generating a partial seizure this method may or may not be able to identify all parts of the network. The reason is that the method is likely to fail if there are too many sources distributed over a spread out region. However it will be able to detect the phase synchronization between two weakly coupled signals by returning a suitable value of *syn* function. But then it will also churn out a high number of false positive. An ROC (receiver operator characteristics) curve analysis of performance of the algorithm will have to be undertaken on a large real data set in order to validate the method for implementation. Determining an appropriate value of the *syn* function for a specific application will remain a tricky issue. ROC curve analysis will be of great use in this regard.

The accuracy of source localization will depend on accuracy of locating the electrodes. For the photogrammetry method for localizing electrode positions the mean error rate is 1.27 mm [22]. Whereas the mean error rate for the electromagnetic method is 1.02 mm, with a mean error of 0.56 mm for calibration objects in both the cases [22]. The photogrammetry method is custom made for the dense array EEG. Dense array EEG is necessary for detecting EEG signals from the source in more ways than what is barely necessary. This is an important advantage for being able to choose multiple collection of four point sets for the same source. The more such blocks of synchronous four points are detected the more will the scope to pinpointedly locate the source, be it a single point or a small region.